\documentclass[runningheads]{llncs}
\usepackage{graphicx}
\usepackage{comment}
\usepackage{amsmath,amssymb} % define this before the line numbering.
\usepackage{color}
\usepackage{breakcites}

\DeclareMathOperator{\EX}{\mathbb{E}}
\DeclareMathOperator{\PP}{\mathbb{P}}

\begin{document}
% \linenumbers
\pagestyle{headings}
\mainmatter
\def\ECCVSubNumber{7}  % Insert your submission number here

\title{Mapping Low-Resolution Images To Multiple High-Resolution Images Using Non-Adversarial Mapping} % Replace with your title

%\begin{comment}
% \titlerunning{ECCV-20 submission ID \ECCVSubNumber} 
% \authorrunning{ECCV-20 submission ID \ECCVSubNumber} 
% \author{Anonymous ECCV submission}
% \institute{Paper ID \ECCVSubNumber}
%\end{comment}
%******************

%\begin{comment}
\titlerunning{Super-Resolution Non-Adversarial Mapping}
\author{Vasileios Lioutas}
\authorrunning{V. Lioutas et al.}

\institute{School of Computer Science\\Carleton University, Ottawa, Canada\\
\email{contact@vlioutas.com}}
%\end{comment}
%******************
\maketitle

\begin{abstract}
Several methods have recently been proposed for the Single Image Super-Resolution (SISR) problem. The current methods assume that a single low-resolution image can only yield a single high-resolution image. In addition, all of these methods use low-resolution images that were artificially generated through simple bilinear down-sampling. We argue that, first and foremost, the problem of SISR is an one-to-many mapping problem between the low-resolution and all possible candidate high-resolution images and we address the challenging task of learning how to realistically degrade and down-sample high-resolution images. To circumvent this problem, we propose SR-NAM which utilizes the Non-Adversarial Mapping (NAM) technique. Furthermore, we propose a degradation model that learns how to transform high-resolution images to low-resolution images that resemble realistically taken low-resolution photos. Finally, some qualitative results for the proposed method along with the weaknesses of SR-NAM are included.
\end{abstract}

\section{Introduction}
The Single Image Super-Resolution (SISR), a technique for restoring a visually pleasing high-resolution (HR) image from its low-resolution (LR) version, is still a challenging task within the computer vision research community \cite{DBLP:journals/corr/CaballeroLAATWS16,DBLP:journals/corr/DongLHT15,Kappeler2016VideoSW,DBLP:journals/corr/KimLL15b,DBLP:journals/corr/LedigTHCATTWS16,8237536,DBLP:journals/corr/SajjadiSH16,DBLP:journals/corr/ShiCHTABRW16,DBLP:journals/corr/TaoGLWJ17}. Since multiple solutions exist for the mapping from LR to HR space, SISR is highly ill-posed and a variety of algorithms, especially the current leading learning-based methods are proposed to address this problem.

Understanding what the SISR problem represents is crucial in order to develop a method that is capable of solving it.  Having a low-resolution image at inference time means that there is no ground truth answer on how the high-resolution counterpart image is generated. That being said, in order to recover a higher-resolution image, assumptions need to be made that do not violate the visible artifacts taken from the low-resolution image. The fine details added to the higher-resolution image are subjective, since they only need to follow certain already visible artifacts from the low-resolution image. The task in SISR is to find a model that learns how to make these assumptions and generate high-resolution images as plausible as possible according to the specific task that is being undertaken like, Face SISR. To this day, all current solutions for the SISR problem attempt to reconstruct a single high-resolution image based on a given low-resolution input image. In other words, the process of generating a high-resolution image is deterministic and given the same low-resolution image multiple times as input will yield the same high-resolution image.

In this paper, we argue that a method for solving the SISR problem should yield multiple high-resolution candidates for the same low-resolution image and we propose an approach to solve this problem. Specifically, the proposed SR-NAM method is an unsupervised method of mapping high-resolutions images to a given low-resolution image. The advantage of this method over others is that it is fast and it requires to optimize only a single representation. This representation attempts to match pre-trained fixed knowledge of both the high-resolution image space as well as the degradation method. To the best of our knowledge, all previous works on SISR, degraded the high-resolution images artificially using down-sampling methods such as bilinear and bicubic algorithms in order to create a dataset of high-resolution images and the associated low-resolution image. Usually, these methods do not perform well when they are used with real-world low-resolution images as shown in \cite{DBLP:journals/corr/abs-1712-06087,DBLP:journals/corr/abs-1807-11458}. In contrast to these approaches, following the work from \cite{DBLP:journals/corr/abs-1807-11458} we propose to use a degradation model to generate a low-resolution image from a high-resolution image that visually appears to be taken from a low quality camera.

\section{Related Work}
\subsection{Image Super-Resolution} The problem of SISR has been widely studied. Early approaches either rely on natural image statistics \cite{5396341,Zhang2010NonLocalKR} or predefined models \cite{Irani:1991:IRI:108693.108696,Fattal:2007:IUV:1276377.1276496,4587659}. Later, mapping functions between LR images and HR images are investigated, such as sparse coding based SR methods \cite{Zeyde:2010:SIS:2187598.2187645,5466111}.

Recently, deep convolution neural networks (CNN) have been shown to be powerful and capable of improving the quality of SR results \cite{zhang2019deep,DBLP:journals/corr/abs-1803-08664,Park_2018_ECCV,wang2018sftgan,8237776}. It needs to be highlighted, that all the aforementioned image super-resolution methods can be applied to all types of images and hence do not incorporate face-specific information, as proposed in our work.

\subsubsection{Face Super-Resolution.} There are many works in the literature focusing specifically on applying SISR techniques to face images. The recent works of \cite{Yu2016UltraResolvingFI,Yu_2018_CVPR,DBLP:journals/corr/abs-1807-11458,bulat2017far} use a GAN-based approach. Other works like \cite{DBLP:journals/corr/abs-1708-03132} used Reinforcement Learning to learn to progressively attend on specific parts of a face image in order to restore them in a sequence procedure. Some other methods \cite{DBLP:journals/corr/abs-1711-10703} work with introducing facial prior knowledge that could be leveraged for better super-resolving face images. The method of \cite{DBLP:journals/corr/ZhuLLT16} performs super-resolution and dense landmark localization in an alternating manner which is shown to improve the quality of the super-resolved faces.

\subsection{Unsupervised domain alignment} Due to the rise of generative adversarial networks (GANs), unsupervised translation across different domains began to generate strong results. Most of the state-of-the-art unsupervised translation methods employ GAN techniques. The most popular extension to the traditional GAN approach is the use of the cycle-consistency which enforces the generated samples mapped between  the two domains to be the same. This approach is widely used by DiscoGAN \cite{DBLP:journals/corr/KimCKLK17}, CycleGAN \cite{CycleGAN2017} and DualGAN \cite{DBLP:journals/corr/YiZTG17}. Recently, StarGAN \cite{DBLP:journals/corr/abs-1711-09020} extended the approach to more than two domains. Our work is built upon Non-Adversarial Mapping (NAM) method \cite{DBLP:journals/corr/abs-1806-00804} and the details are described on Section \ref{sec:srnam}.

\section{Super-Resolution using NAM}
As mentioned in Section 1, we propose two main models: the degradation model and the Super-Resolution NAM model. The degradation model is designed to take as input a HR image and produce an output resembles a realistically taken LR image. The SR-NAM model then uses a pre-trained face generator and the degradation model to learn to infer with no supervision the predicted HR image.

\subsection{Datasets}
\label{sec:datasets}
This section describes the HR and LR datasets used during training and testing. In order to train the degradation model, a dataset with real-world LR images is needed. Searching  the literature, the only available dataset that fulfills the requirements is the one described in \cite{DBLP:journals/corr/abs-1807-11458}. Thus, we decided to contact the authors in order to get access to the exact subset of data that they used in their work.

\subsubsection{HR dataset.} Following \cite{DBLP:journals/corr/abs-1807-11458}, the High-Resolution (HR) image dataset is composed of 182,866 face images of size 64$\times$64. The authors of the dataset aimed to create a dataset that is as balanced as possible in terms of facial poses. The dataset is a combination of subsets of 4 popular face datasets. Specifically, from Celeb-A \cite{DBLP:journals/corr/LiuLWT14}, they randomly selected 60,000 faces. Additionaly, they used the whole AFLW \cite{6130513} dataset. Finally, a subset of the LS3D-W \cite{bulat2017far} and VGGFace2 \cite{DBLP:journals/corr/abs-1710-08092} datasets is been used. The dataset includes many face images of various poses, illuminations, expressions and occlusions. 

\subsubsection{LR dataset.} The authors in \cite{DBLP:journals/corr/abs-1807-11458} created a real-world Low-Resolution (LR) image dataset from the Widerface \cite{yang2016wider} face dataset. This dataset is very large in scale and diverse in terms of faces and it contains real-world taken pictures with various forms of noise and degradation. The dataset is composed of 50,000 images of size 16$\times$16 where 3,000 randomly selected and kept for testing.

\begin{figure*}[t]
    \centering
    \includegraphics[width=\textwidth]{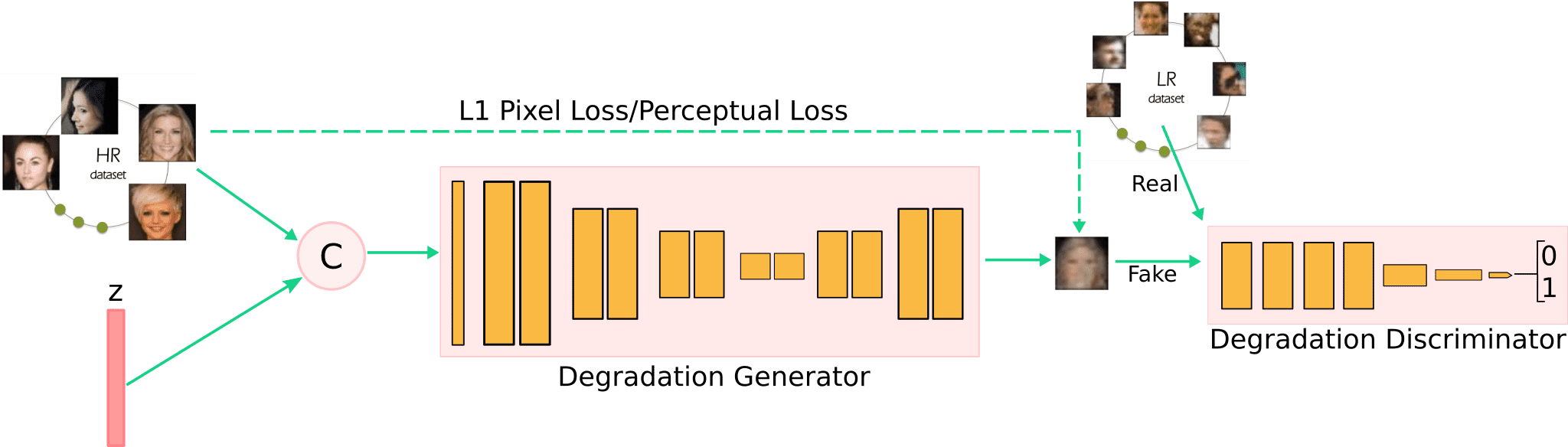}
    \caption{Overall proposed architecture and training pipeline for the degradation model.}
    \label{fig:degradation_model}
\end{figure*}

\subsection{Degradation Model}
The degradation model is inspired by \cite{DBLP:journals/corr/abs-1807-11458}. The overall architecture is composed by a generator and a discriminator network. Both models are based on the ResNet architecture \cite{DBLP:journals/corr/HeZRS15}. The overall architecture and training procedure is shown on Figure \ref{fig:degradation_model}.

\subsubsection{Degradation Generator.} A HR image coming from the HR dataset is used as input to the degradation generator. The architecture is similar to the one used in \cite{DBLP:journals/corr/abs-1807-11458}. The network is following an encoder-decoder schema and it composed by 12 residual blocks equally distributed in 6 groups. The resolution is dropped 4 times using a pooling layer. Specifically, given the input size of 64$\times$64, we degrade it to 4$\times$4 px. Next, the size is increased twice to 16$\times$16 using a pixel shuffle layer \cite{Shi_2016}.

In addition to the HR image, a noise vector is concatenated that was projected and then reshaped using a fully connected layer in order to have the same size as one image channel. The intuition behind this is that degrading a HR image to LR image is an one-to-many problem where an HR image can have multiple corresponding LR images.

\subsubsection{Degradation Discriminator.} The discriminator is similar to the ResNet architecture and it consists of 6 residual blocks without any batch normalization in between, followed by a fully connected layer. To drop the resolution of the 16$\times$16 image, max-pooling is been used for the last two blocks.

\subsubsection{Degradation Loss.} The degradation generator and discriminator networks were trained with a total loss which is a combination of a GAN loss and a pixel loss defined as
\begin{equation}
\label{eq:total_loss}
    l = {\alpha}l_{pixel} + {\beta}l_{GAN}
\end{equation}
where $\alpha$ and $\beta$ are the corresponding weights.

Following \cite{DBLP:journals/corr/abs-1807-11458}, we used the GAN loss defined as
\begin{equation}
\label{eq:gan_loss}
    l_{GAN} = \underset{x \sim \PP_r}{\EX}[min(0,-1+D(x))]+\underset{\hat{x} \sim \PP_g}{\EX}[min(0,-1-D(\hat{x}))]
\end{equation}
where $\PP_r$ is the LR data distribution and $\PP_g$ is the generator $G$ distribution defined by $\hat{x} = G(x)$. For the GAN loss, an ``unpaired'' training setting is used where the real-world images from the LR dataset are enforcing the output of the generator (whose input is images from the HR dataset) to be contaminated with real-world noisy artifacts. According to \cite{pmlr-v70-arjovsky17a}, using Wasserstein distance as GAN loss, greatly improves the stability of the GAN model. In \cite{pmlr-v70-arjovsky17a}, in order to enforce the Lipschitz constraint the authors used weight clipping. We decided to enforce this constraint using the more recent and improved approach of gradient penalty as described in \cite{DBLP:journals/corr/GulrajaniAADC17}.

Finally, the $l_{pixel}$ loss is used to enforce the output of the generator to have similar content (i.e. face identity, pose and expression) with the original HR image defined as
\begin{equation}
\label{eq:pixel_loss}
    l_{pixel} = {\gamma}l_1 + {\delta}l_{VGG}
\end{equation}
where $\gamma$ and $\delta$ are the corresponding weights. The $l_1$ loss is defined as
\begin{equation}
\label{eq:num1}
l_1 = \sum_{y \in \text{HR}} \|F(G(I_y^{hr})), I_y^{hr}\|_1
\end{equation}
where $F$ is an up-scaling function. Also, we decided to use the perceptual loss \cite{DBLP:journals/corr/JohnsonAL16} which was found to give perceptually pleasing results. This is defined as
\begin{equation}
\label{eq:perceptual}
l_{VGG}  = \sum_{y \in \text{HR}} \sum_i \|\phi_i(F(G(I_y^{hr}))), \phi_i(I_y^{hr})\|_1
\end{equation}
where $\phi_i()$ be the features extracted from a deep-network at the end of the $i$-th block (we use VGG \cite{7486599}).

\subsection{HR Generative Model}
\label{sec:gen_theory}
\begin{figure*}[t]
    \centering
    \includegraphics[width=\textwidth]{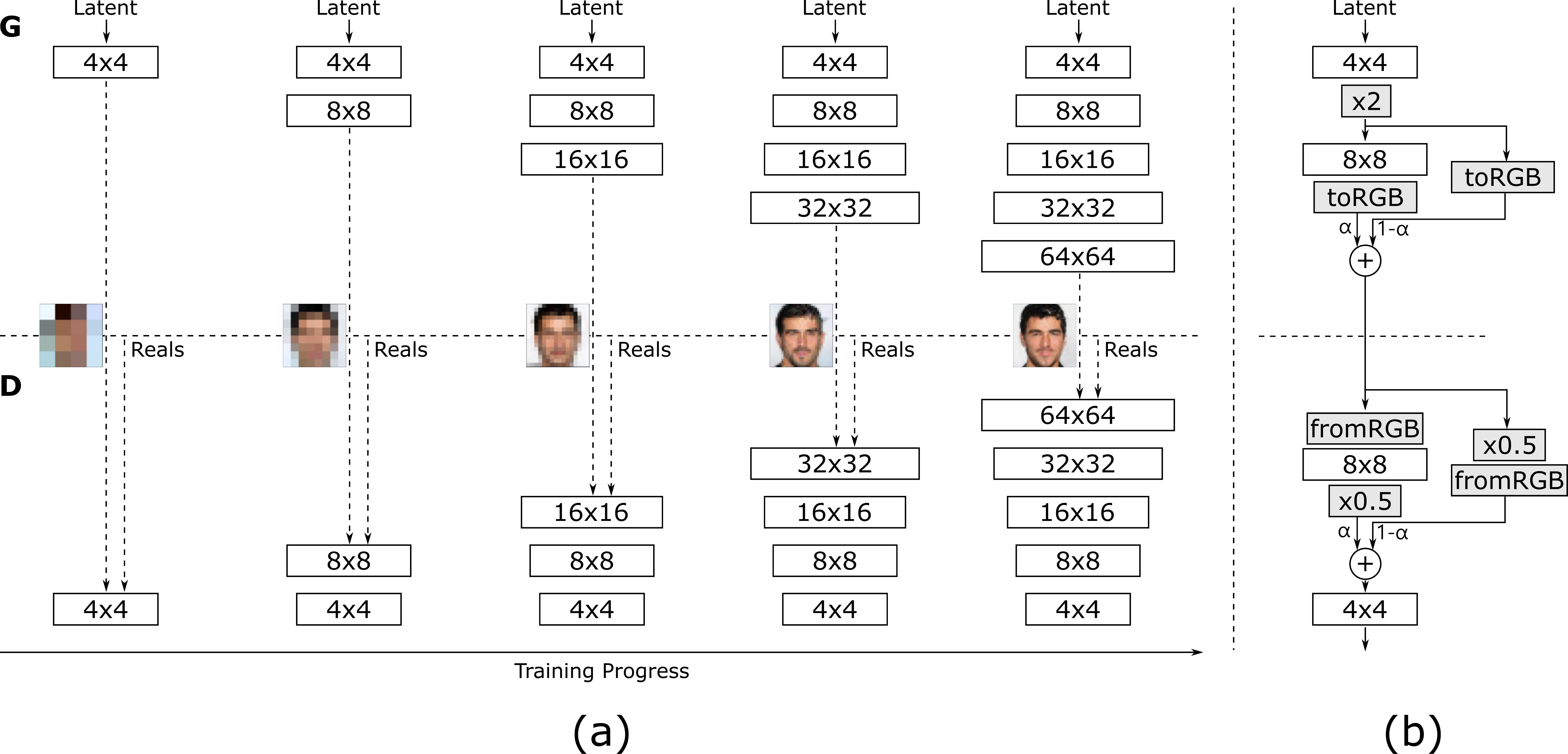}
    \caption{(a) The Progressive GAN training process. During training both the generator G and the discriminator D increase the lower spatial resolution to a two times higher resolution progressively, keeping the existing layers trainable throughout the process. (b) During the upscaling, the higher resolution layers act as residual blocks. The \framebox{\small toRGB} layer projects feature vectors to RGB colors and \framebox{\small fromRGB} does the reverse. Figure adapted from \cite{DBLP:journals/corr/abs-1710-10196}.}
    \label{fig:progan_model}
\end{figure*}

This section describes the HR generator that is used to generate HR images given a latent representation. Pre-training a high-quality and generalized face generator is crucial for the success of the SR-NAM model. For this reason, we decided to experiment with the Progressive GAN architecture as described in \cite{DBLP:journals/corr/abs-1710-10196}. The authors of that paper showed that their model is very effective on generating good quality HR images given enough face images. The overall architecture can be seen on Figure \ref{fig:progan_model}.

\subsubsection{Progressive GAN.} Following \cite{DBLP:journals/corr/abs-1710-10196}, the idea behind the progressive generator architecture is to start with a low-resolution image, and then progressively increase the resolution by adding layers to the network. This is illustrated on Figure \ref{fig:progan_model}. This incremental procedure helps the training to first, find the large-scale structure of the image distribution and then to shift attention to progressively finer scale details, instead of trying to learn everything simultaneously. The discriminator and generator networks are mirrored and grow in a synchronous way. All the previous layers up to the new resolution remain trainable throughout the training process. The new layers that are added to the network are faded in smoothly to avoid sudden changes to the already well-trained lower resolution layers (Figure \ref{fig:progan_model}).

\subsection{SR-NAM}
\label{sec:srnam}
This section describes the Super-Resolution using Non-Adversarial Mapping approach that is used to retrieve multiple HR images from a single LR image, which is the main focus of this work. Let $I^{lr}$ be the low-resolution space and $I^{hr}$ be a high-resolution space, consisting of two set of images $\{I_i^{lr}\}$ and $\{I_i^{hr}\}$ respectively. The objective is to find every image $I_i^{hr}$ in the high-resolution space, that is analogous to an image $I_i^{lr}$ in the low-resolution domain. Each $I_i^{hr}$ should visually appear as an image from the high-resolution space but preserve the unique content of the original $I_i^{lr}$ image.

\subsubsection{Non-Adversarial Mapping.} NAM \cite{DBLP:journals/corr/abs-1806-00804} is a method for unsupervised mapping across image domains. For using this approach, you must have a pre-trained unconditional model $G(z)$ of the $X$ domain, where $X$ in our case is the $I^{hr}$ high-resolution space. In addition, you must have a set of $Y$ domain training images $\{y\}$, which in our case corresponds to $\{I_i^{lr}\}$ from the low-resolution space.

\begin{figure*}[t]
    \centering
    \includegraphics[width=\textwidth]{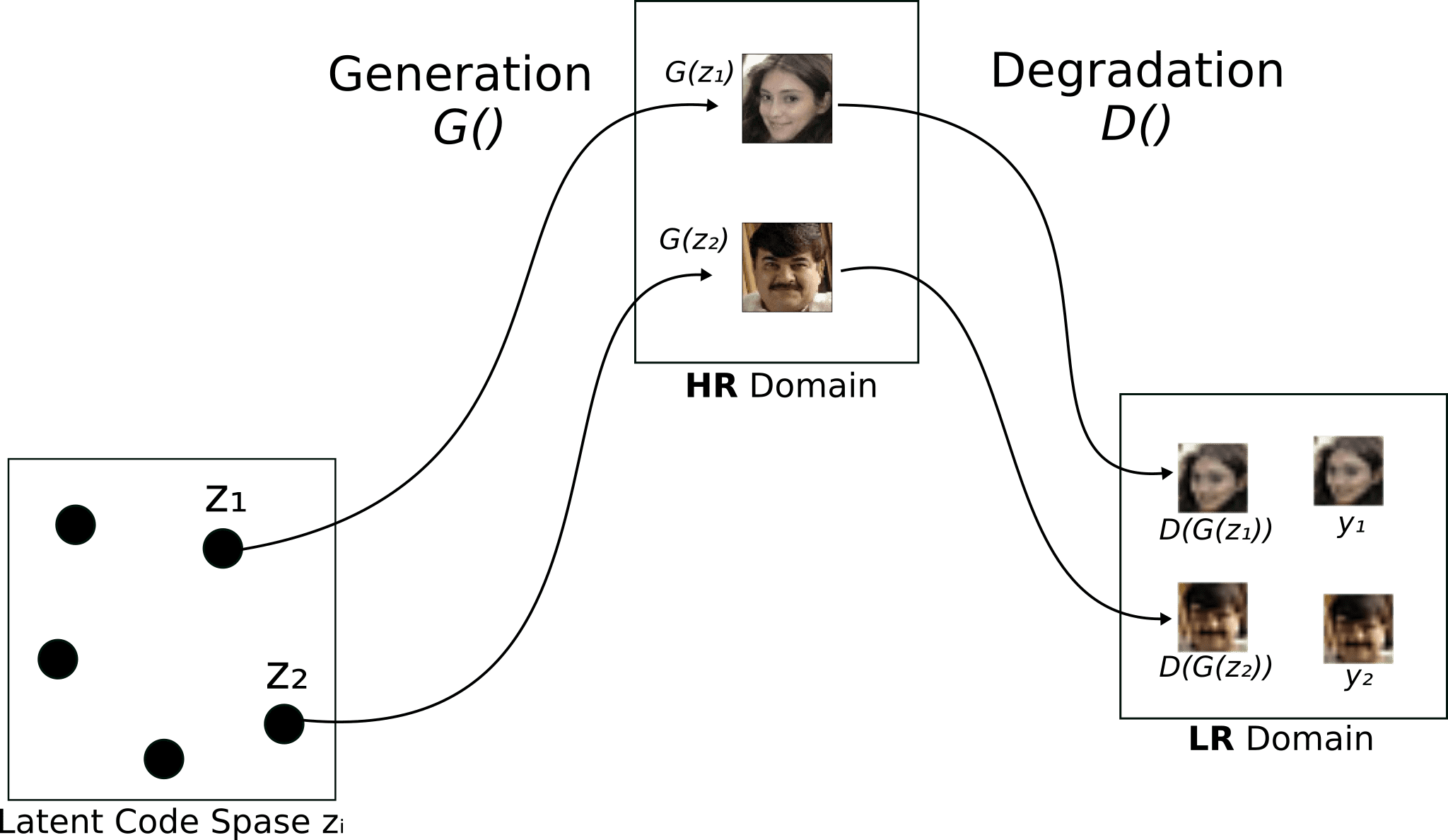}
    \caption{Given a HR generator $G$ and training samples $\{I_i^{lr}\}$, SR-NAM jointly learns the degradation network $D : I^{hr} \rightarrow I^{lr}$ and the latent vectors {$z_i$} that give rise to samples {$D(G(z_i))$} that resemble the training images in $I^{lr}$.}
    \label{fig:srnam_model}
\end{figure*}

Given a pre-trained, high-resolution domain, generative model $G(z)$, a pre-trained degradation model $D(I_i^{hr})$ and a set of $\{I_i^{lr}\}$ training images, NAM estimates the latent code $z_{I_i^{lr}}$ for every $I^{lr}$ training image so that the generated $I_i^{hr}$ image from the latent code $D(I_i^{hr})$ maps to the low-resolution image $I_i^{lr}$. Figure \ref{fig:srnam_model} shows exactly this process. The entire optimization problem is defined as
\begin{equation}
\label{eq:nam_optim}
    argmin_{z_{I_i^{lr}}} \sum_{I_i^{lr} \in I^{lr}} \|D(G(z_{I_i^{lr}})), I_i^{lr}\|_1
\end{equation}

The advantages of NAM include that it does not use adversarial training to learn the mapping between high- and low-resolution images. In addition, the mapping can be applied to many situations and multiple solutions can be recovered for a single low-resolution input image. NAM is also capable of using a pre-trained high-resolution model as well as a pre-trained low-resolution model that only need to be estimated once.

In contrast to \cite{DBLP:journals/corr/abs-1806-00804}, we decided not to include the perceptual loss as part of the optimization objective. This is because, although perceptual loss successfully yields perceptually pleasing results (i.e. following perceptually the content of the low-resolution image such as similar pose, expression, face geometry etc.), it is of little use when the goal is to recover as close as possible a low-resolution image to the higher resolution counterpart since it can produce images that are correct perceptually but totally different visually. Thus, we decided to only minimize the $L_1$ loss between $D(G(z_{I_i^{lr}}))$ and $I_i^{lr}$.

\subsubsection{Inference.} Since all the networks inside the SR-NAM model are already pre-trained and fixed, only the latent code $z_{I_i^{lr}}$ needs to be optimized each time. To infer an analogy of a new $I_i^{lr}$ image, we need to recover the latent code $z_{I_i^{lr}}$ which would yield the optimal reconstruction. The generated $I_i^{hr}=G(I_i^{lr})$ high-resolution image is the proposed solution to the low-resolution image $I_i^{lr}$.

\subsubsection{Multiple Solutions.} In order to produce multiple HR images from a LR image, it is sufficient to sample a different randomly initialized latent code $z_{I_i^{lr}}$. This is because the problem space is non-convex, thus starting from a different point in the space can yield different final analogies.

\section{Experiments}
In this section, we demonstrate the effectiveness of the SR-NAM approach by reporting some qualitative results on both the HR as well as the LR dataset. Further, we show the performance of both the degradation network and the progressive GAN procedure.

\subsection{Implementation Details}
In this section, we give a detailed description of the procedure used to generate the experimental results presented in this project.

\begin{figure*}[t]
    \centering
    \includegraphics[width=\textwidth]{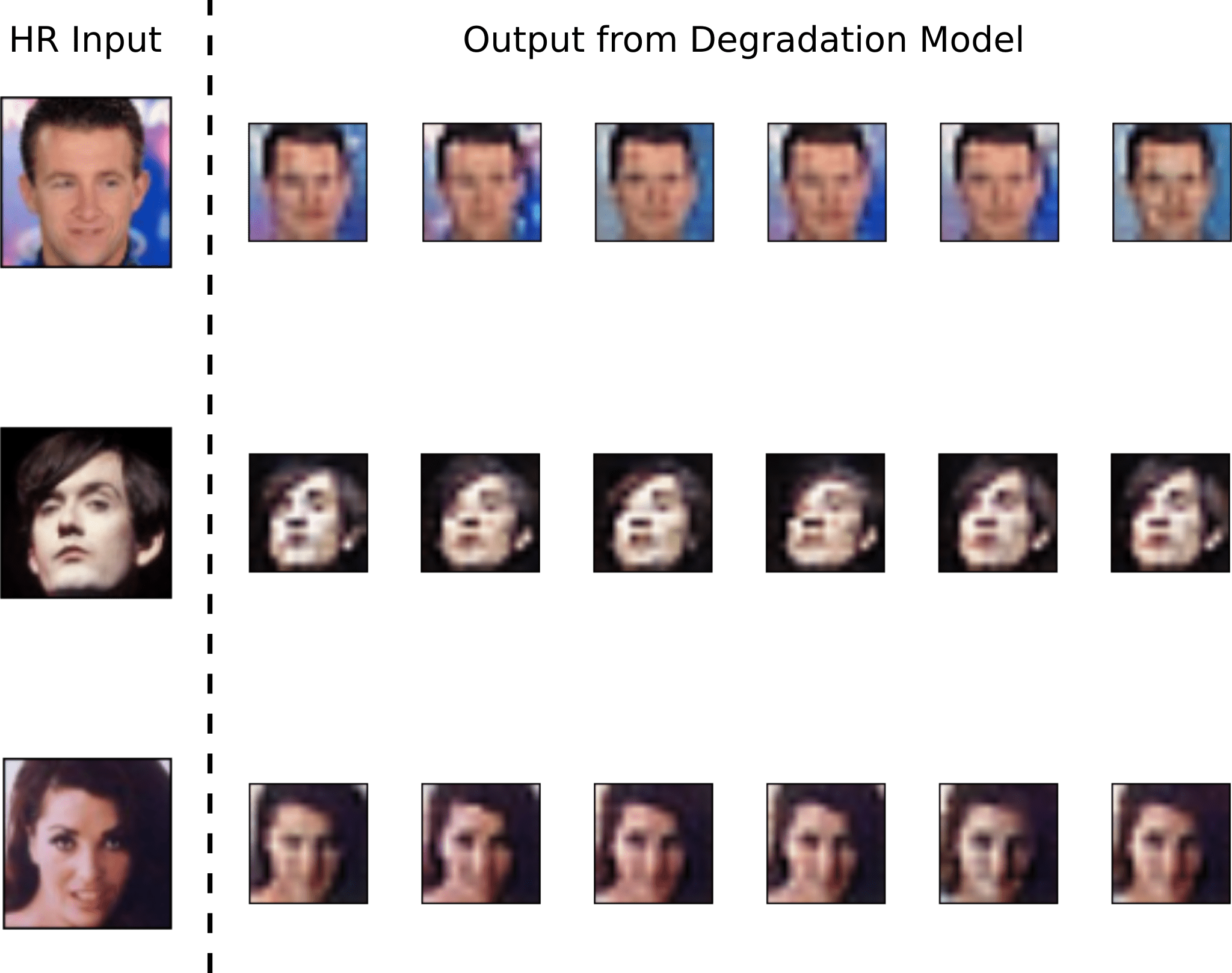}
    \caption{Examples of different low-resolution samples produced by the degradation network for different input noise vectors.}
    \label{fig:degra_results}
\end{figure*}

\subsubsection{Degradation Model.} The intuition behind the degradation model is to create a model that can generate a realistically taken LR face image based on a HR image. This model is used both to create the ground truth LR images of the HR image dataset as well as the degradation model for the SR-NAM approach, which is responsible to convert the generated HR candidate back to the LR image in order to be compared with the ground truth LR image. We trained the model for 500,000 iterations and we used the Adam\cite{DBLP:journals/corr/KingmaB14} optimizer with default settings. The discriminator was set to be trained for 5 more iterations on each iteration step before training the generator. The $\lambda$ value for gradient penalty was set to 10. The latent size for the input noise was set to be 100. In addition, the batch size was chosen to be 64. Finally, a pre-trained version of the VGG network with 19 layers has been used for the perceptual loss.

\subsubsection{HR Generative Model.} SR-NAM takes as input a pre-trained generative model of the HR image domain. As mentioned in Section \ref{sec:gen_theory}, we decided to use the ProGAN model. The depth of the resolution was set to 5 which translates to 64$\times$64 images. Each resolution was trained for 10, 20, 20, 20 and 50 epochs respectively. The batch sizes were also set to 64, 64, 64, 32 and 16 for each resolution. Due to the complexity of the face generation problem, we decided to use a latent size of 512. In addition, we used Adam\cite{DBLP:journals/corr/KingmaB14} optimizer with default settings for the optimization procedure. Finally, the training of the model took approximately two weeks to complete in a single NVIDIA GeForce GTX 1080 Ti GPU card.

\subsubsection{SR-NAM Model.} Since SR-NAM takes as input both the HR generative models and the degradation model as pre-trained and fixed networks, the only optimization that is needed to be done is on the latent codes for each training/testing example. Again, we chose to use the Adam optimizer with default settings. Since the results are sensitive to how much generalized is the HR generative model, the number of iterations that each example needs in order to find and recover a corresponding HR image from the learned HR space varies. Empirically, we found a good number of iterations to be between 250 and 500 iterations.

\subsection{Degradation Model Results}
Figure \ref{fig:degra_results} shows the results of the trained degradation model. It is clear that the model is able to produce a 16$\times$16 low-resolution image given a 64$\times$64 high-resolution image. It is worth noting that the network can model a variety of image degradation styles, in different levels, such as blurriness, distortion, colouring, illumination, face geometry etc. Thus, it learns the different types of noise that is more probably to be produced in a real-world setting and as like the image was taken using a low quality camera.

\begin{figure}
    \centering
    \includegraphics[width=0.8\textwidth]{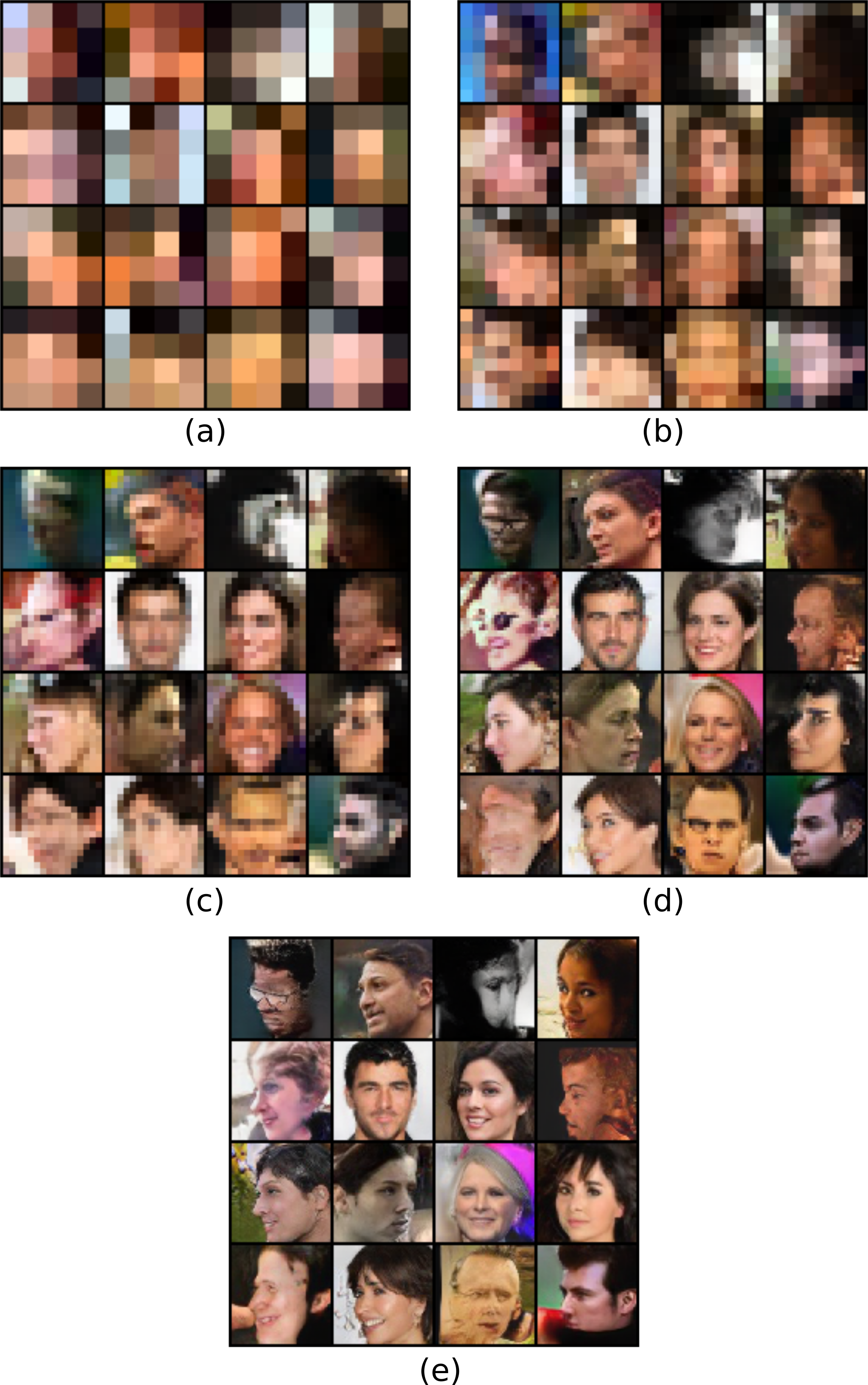}
    \caption{Qualitative results showing the effectiveness of the progressive generator for five different resolutions: 4$\times$4, 8$\times$8, 16$\times$16, 32$\times$32 and 64$\times$64. All the examples were generated using a fixed random noise input.}
    \label{fig:progan_results}
\end{figure}

\subsection{Progressive Generator Results}
We show examples of a variety of face images generated at 64$\times$64 by ProGAN. Figure \ref{fig:progan_results} shows generated results of faces at 4$\times$4, 8$\times$8, 16$\times$16, 32$\times$32 and finally at 64$\times$64 resolution, using a fixed random noise, given as input on each resolution. The progressive generator successfully learns to produce clear 64$\times$64 face images. 

\subsection{SR-NAM Results}
In this section, we evaluate the performance of SR-NAM. The details of our experiments are as follows:

\subsubsection{Performance metrics.} The scope of this paper is to create an approach capable of generating multiple HR face images that correspond to a LR input face image. To date, based on our knowledge, there is no quantitative metric that can be used to measure if an image follows both perceptually and visually a given ground truth image. To overcome this issue, we propose the following new metric that can be used to measure the performance of each different generated HR image. This metric will use a facial landmark localization algorithm such as \cite{bulat2017far}, to find all the facial landmarks from the LR image and compare them with the landmarks from the original HR image. The metric is defined as
\begin{equation}
    \text{heatmap\_metric} = \frac{1}{N}\sum_{n=1}^N \sum_{ij} (\hat{M}_{i,j}^n - M_{i,j}^n)^2
\end{equation}
where $\hat{M}_{i,j}^n$ is the heatmap corresponding to the $n$-th landmark at pixel $(i,j)$ produced by the facial landmark localization algorithm with input the generated HR image $\hat{I}_{HR}$. $M_{i,j}^n$ is the heatmap obtained by running the algorithm on the original HR image $I_{HR}$. Due to computational constrains, we did not perform quantitative evaluation on the proposed approach using this metric but it is worth noting that this can be a possible new metric to the problem.

\begin{figure*}[t]
    \centering
    \includegraphics[width=\textwidth]{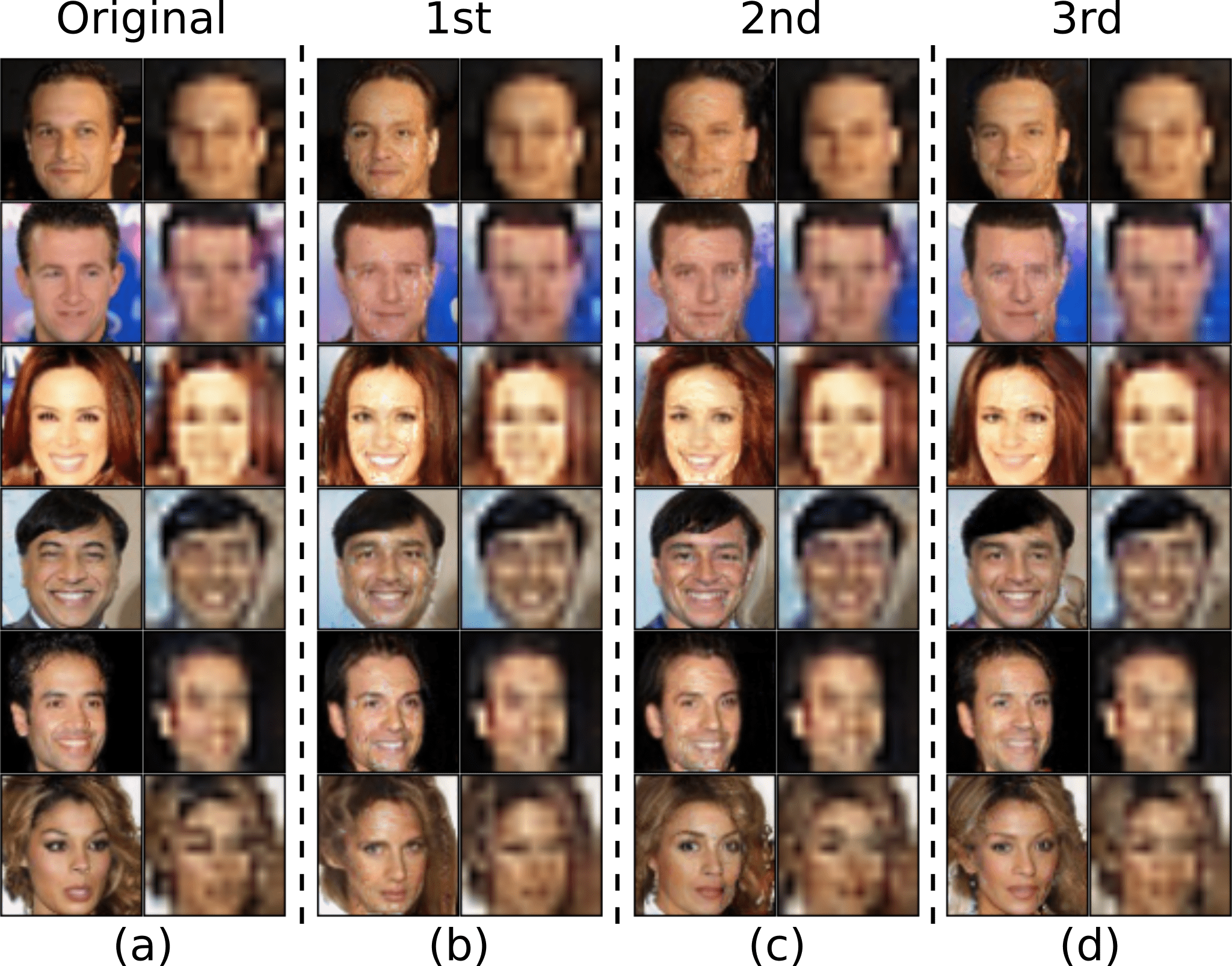}
    \caption{Results of the SR-NAM method on the HR dataset that described on Section \ref{sec:datasets}. The first set of columns shows the original HR image and the degraded corresponding image after using the pre-trained degradation model. The rest three set of columns show the multiple generated HR images with different random initialization each time along with the associated LR image coming from the generated HR image.}
    \label{fig:sr_nam_results_good}
\end{figure*}

For evaluating the quality of the generated HR images, in literature there are mainly two standard metrics that have been used, namely the PSNT and SSIM \cite{1284395} metrics. Up to date, those metrics are heavily criticized by the research community \cite{DBLP:journals/corr/LedigTHCATTWS16,DBLP:journals/corr/abs-1712-02765} as they fail to perceive the real image quality and are considered poor measures. Since, the scope of the project is not to produce better quality HR images but to find a way of producing multiple HR corresponding images from a LR input image, there was no reason for computing these metrics.

\begin{figure}
    \centering
    \includegraphics[width=\textwidth]{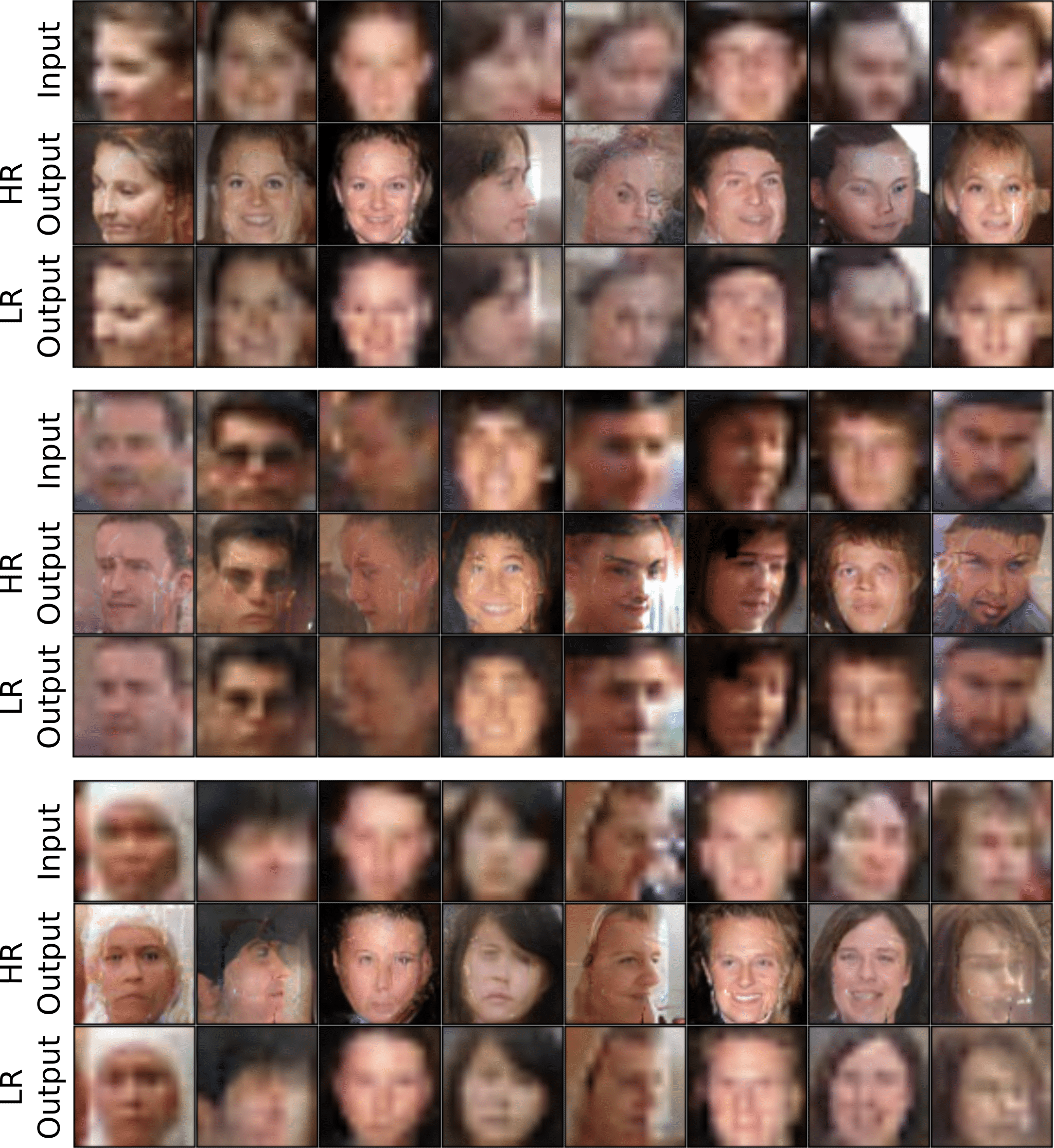}
    \caption{SR-NAM results on the real-world LR dataset. Top row on each set of examples shows the LR input image, middle row shows the generated HR output image and bottom row shows the degradation of the generated HR output that is used to match the LR input image.}
    \label{fig:lr_nam_results_good}
\end{figure}

\subsubsection{Qualitative Results.} Figure \ref{fig:sr_nam_results_good} shows qualitative results for several images from the HR dataset. The first set of columns shows the original HR image and the corresponding degraded LR image. The rest set of columns show the generated HR image and the corresponding degraded LR image on three different reconstructions using different random initializations on the $z_{I_i^{lr}}$ each time. The proposed approach is unsupervised i.e. on inference time, the $z_{I_i^{lr}}$ needs to be learned using the objective of comparing the input LR image with the degraded HR image until they match. Thus, it is worth noting that the model successfully learns to match the input LR image with the generated LR images as it is visualized on Figure \ref{fig:sr_nam_results_good}. In addition, it successfully generates new plausible reconstructions of the input LR image each time.

We also show results on Figure \ref{fig:lr_nam_results_good} using the LR dataset that described in Section \ref{sec:datasets}. It is clear that it successfully reconstructs a more clear HR image compared to the LR image used as input. The faces have still a lot of noise and not all the times follow exactly the face geometry and all the other artifacts but still closely resembles the LR input image. 

\begin{figure*}[t]
    \centering
    \includegraphics[width=\textwidth]{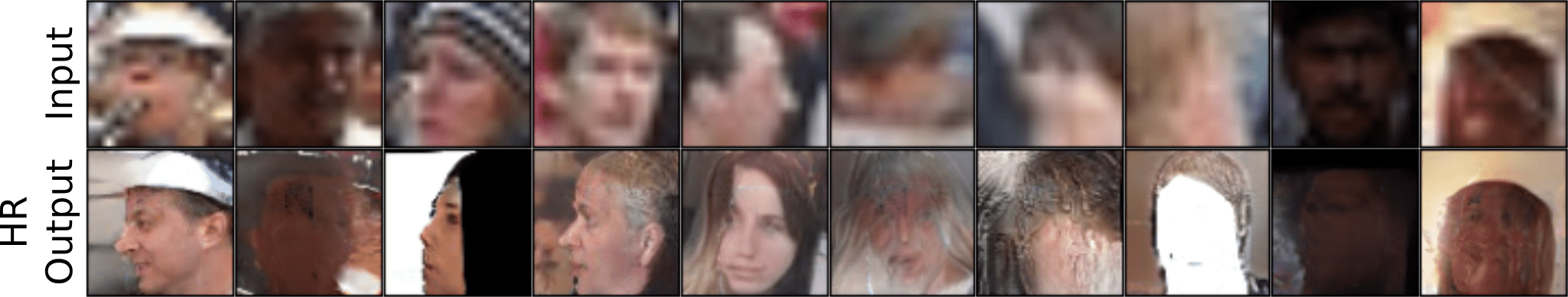}
    \caption{SR-NAM failure cases on the real-world LR dataset. Top row shows the input LR image and bottom row shows the generated HR image.}
    \label{fig:lr_nam_results_bad}
\end{figure*}

\subsection{Failure Cases and Discussion}
The success of this approach lies in obtaining a very well trained and generalized HR Generative model. Without having a generator that can simulate practically all the possible face generations, this method would not work in practice. By no-means, we claim that the current training procedure that we performed is enough to solve this problem but as a proof of concept, it is clear that given the appropriately trained and generalized models, this method can yield different reconstructions each time. In Figure \ref{fig:lr_nam_results_bad}, we demonstrate some of the failure cases where the face either did not resemble the LR face or was not successful on reconstructing a face at all. Furthermore, Figure \ref{fig:lr_nam_results_bad} depicts some cases where reconstructing a HR face can be challenging due to distortion and illumination.

\section{Conclusion and Future Work}
In this paper, we presented a method for face super-resolution which does not assume that there is only a single HR image from a LR input image, but rather maps the LR image to multiple candidate HR images. In addition, the presented method does not assume as input an artificially generated LR image but aims to produce results applied to real-world LR images. We discussed the advantages and disadvantages of the presented method, including that the power of the method lies mainly on the training of the HR generator and that this model needs to be well generalized in all possible faces in order to perform well given new unseen examples. Finally, we demonstrated some qualitative results of the SR-NAM method on both real-world low-resolution images and degraded high-resolution images.

%\clearpage
\bibliographystyle{splncs04}
\bibliography{egbib}
\end{document}